\documentclass[floatfix,aps,twocolumn,prb,superscriptaddress, ]{revtex4}
\usepackage{amsmath}
\usepackage{amssymb}
\usepackage{graphicx}
\usepackage{dcolumn}
\usepackage{natbib}
\usepackage{bm}
\usepackage{verbatim}

\begin{document}

\title{Optical characterization of Bi$_2$Se$_3$ in a magnetic field: infrared evidence for magnetoelectric coupling in a topological insulator material}
\author{A. D. LaForge}
\email{alaforge@ucsc.edu}
\affiliation{Department of Physics, University of California, San Diego, La Jolla, California 92093, USA}
\affiliation{Department of Physics, University of California, Santa Cruz, Santa Cruz,
California 95064, USA}
\author{A. Frenzel}
\altaffiliation{Present address: Department of Physics, Harvard University, Cambridge, Massachusetts 02138, USA}
\affiliation{Department of Physics, University of California, San Diego, La Jolla, California 92093, USA}
\author{B. C. Pursley}
\affiliation{Department of Physics, University of California, San Diego, La Jolla,
California 92093, USA}
\author{Tao Lin}
\affiliation{Department of Physics, University of California, Riverside, Riverside,
California 92521, USA}
\author{Xinfei Liu}
\affiliation{Department of Physics, University of California, Riverside, Riverside,
California 92521, USA}
\author{Jing Shi}
\affiliation{Department of Physics, University of California, Riverside, Riverside,
California 92521, USA}
\author{D. N.~Basov}
\affiliation{Department of Physics, University of California, San Diego, La Jolla,
California 92093, USA}
\date{\today }

\begin{abstract}
We present an infrared magneto-optical study of the highly thermoelectric narrow-gap semiconductor Bi$_2$Se$_3$. Far-infrared and mid-infrared (IR) reflectance and transmission measurements have been performed in magnetic fields oriented both parallel and perpendicular to the trigonal $c$ axis of this layered material, and supplemented with UV-visible ellipsometry to obtain the optical conductivity $\sigma_1(\omega)$. With lowering of temperature we observe narrowing of the Drude conductivity due to reduced quasiparticle scattering, as well as the increase in the absorption edge due to direct electronic transitions. Magnetic fields $H \parallel c$ dramatically renormalize and asymmetrically broaden the strongest far-IR optical phonon, indicating interaction of the phonon with the continuum free-carrier spectrum and significant magnetoelectric coupling. For the perpendicular field orientation, electronic absorption is enhanced, and the plasma edge is slightly shifted to higher energies. In both cases the direct transition energy is softened in magnetic field. 
\end{abstract}

\maketitle

\section{Introduction}

 Since the first synthesis of Bi$_2$Se$_3$ in the late 1950s,\cite{Black-Spencer-Bi2Se3-JPhysChemSol1957} a rich body of theoretical and experimental work has grown out of the effort to explain and exploit the large thermoelectric effect which the material exhibits.\cite{DiSalvo-thermoelectric-Science1999} Recently this compound was vaulted back into the forefront of the condensed matter field after being named a prime candidate for the physical realization of topological surface states.\cite{Zhang-Bi2Se3-Dirac-cone-NatPhys2009} A topological insulator has an energy gap in the bulk but, due to spin-orbit coupling, possesses one or more robust metallic surface states which are protected by time-reversal symmetry. The first three-dimensional material to exhibit this behavior is the alloy Bi$_x$Sb$_{1-x}$.\cite{Hsieh-Hasan-BiSb-ARPES-Nature2008} Recent band structure calculations\cite{Zhang-Bi2Se3-Dirac-cone-NatPhys2009} and experiments,\cite{Xia-Hasan-Bi2Se3-ARPES-NatPhys2009,Hsieh-Hasan-helical-dirac-fermions-Nature2009} however, indicate that Bi$_2$Se$_3$ and several related layered compounds exhibit a single Dirac cone on the Fermi surface, a hallmark of a topological insulator. Beyond the inherent importance of exploring a complex phase of quantum matter, these systems are of great interest for device applications involving quantum computing\cite{Fu-Kane-topo-quantum-computing-PRL2008} and photonics,\cite{Plucinski-optical-fiber-SHG-OpticsComm2002} the latter due to nonlinear electron-phonon interaction effects.

Here we present an infrared spectroscopic study of Bi$_2$Se$_3$ in magnetic field. The original motivation for this work was to search for signatures of the topological insulator state, a proposal with twofold justification. First, the topological nature of such a material is theorized to be sensitive to the application of electric and  magnetic fields,\cite{Qi-Zhang-topological-field-theory-PRB2008,Schnyder-topological-PRB2008} and such a tuning of the surface states may display a spectroscopic signature. Second, the magnetic field allows the probing of band dispersion via the cyclotron resonance. As mentioned above, a defining characteristic of the topological surface state is the existence of an odd number of Dirac cones at the Fermi surface,\cite{Zhang-Bi2Se3-Dirac-cone-NatPhys2009} which in turn prescribes the presence of massless Dirac fermions moving at the Fermi velocity. Such quasiparticles can be detected by angle-resolved photoemission spectroscopy,\cite{Zhou-Lanzara-graphite-dirac-fermions-ARPES-NatPhys2006,Xia-Bi2Se3-ARPES-condmat2008} or alternatively by measurement of the cyclotron resonance. In the latter, optical or tunneling spectroscopy techniques measure transitions between Landau levels in an applied magnetic field and massless quasiparticles are distinguished from massive ones by their square root, rather than linear, dependence of transition energy upon magnetic field.\cite{Li-Andrei-graphite-landau-levels-tunneling-NatPhys2007,Jiang-Henriksen-Kim-graphene-IR-LL-PRL2007} Thus, infrared magnetospectroscopy is an ideal tool for probing the topological insulator quantum state. 

As we will show below, the doping of the sample under investigation was found to place the Fermi surface away from the Dirac cone, partially obscuring the observation of topological surface states.\cite{Analytis-Shen-Bi2Se3-SdH-ARPES-condmat2010,Checkelsky-conductance-PRL2009} However, the remarkable field-induced effects uncovered by this study, including broad transfer of spectral weight  and strong magnetoelectric coupling, demonstrate the intrinsic complexity of the host material and motivate further investigation of topological effects. 

Bismuth selenide (Bi$_2$Se$_3$), a member of the V$_2$VI$_3$  group of materials (V= Bi, Sb, S; VI = Se, Te, S), crystallizes in a rhombohedral structure (point group $\bar{3}mD_3d$).\cite{Black-Spencer-Bi2Se3-JPhysChemSol1957} Five-atom layers, known as quintuple layers, are oriented perpendicular to a trigonal $c$ axis and the covalent bonding within each quintuple layer is much stronger than weak van der Waals forces bonding neighboring layers. Due to selenium vacancies the material is easily $n$ doped over a wide range of carrier concentrations,\cite{Kohler-Fischer-Bi2Se3-high-carrier-PhysSolStat1975,Navratil-Se-vacancies-JSolStatChem2004} but has recently been $p$ doped as well.\cite{Hor-Bi2Se3-p-type-arxiv2009} Transport and optical experiments\cite{Black-Spencer-Bi2Se3-JPhysChemSol1957} have determined the semiconducting gap to be  approximately 0.25-0.35 eV, in good agreement with theoretical calculations.\cite{Zhang-Bi2Se3-Dirac-cone-NatPhys2009} Further studies have investigated the details of the interband transitions\cite{Black-Spencer-Bi2Se3-JPhysChemSol1957,Greenaway-Harbeke-Bi2Se3-band-structure-JPCS1965} and phonons,\cite{Richter-Kohler-Becker-Bi2Se3-phonons-PhysSolStat1977} as well as characterized the doping trends with the substitution of Te, Sb,\cite{Kulbachinskii-BiSbSe-SdH-PRB} As,\cite{Sklenar-BiSeAs-optics-CrystResTech2000} Fe,\cite{Kulbachinskii-BiFeSe-SdH-JMMM2004} Mn,\cite{Janicek-BiMnSe-PhysicaB2008} and other elements.

\section{Magneto-optical experiment}

\subsection{Sample information}

Single crystals of Bi$_2$Se$_3$ were prepared by melting stoichiometric alloys of high
purity Bi$_2$Se$_3$ (99.999\%) in a vacuum-sealed quartz tube (with diameter 1/4 or 3/8 in.).
Typically the vacuum of the tube is about $6 \times 10^6$ Torr. The sealed sample was heated to $850^{\circ}$ C and then cooled over a period of three days, from 850 to $650^{\circ}$ C, and annealed at that temperature for a week. The sample was then slowly cooled  to room temperature. Single crystals were obtained and could be easily cleaved from the boule. Transport measurements at room temperature determined a carrier density of $4 \times 10^{18}$ cm$^{-3}$ and resistivity of 1.04 m$\Omega$cm. Reflectance measurements were performed upon a large single crystal of dimensions 3 mm x 4 mm x 2 mm thick. Transmission measurements were performed upon a thin (approximately 40 $\mu$m) flake cleaved from the same crystal.

\subsection{Experimental description}

Infrared reflectance and transmission measurements were performed in a novel magneto-optical apparatus developed at UCSD (see Fig. 1). In this system a translator actuates a helium-flow cryostat through the side port of a  split-coil superconducting magnet, yielding highly repeatable spectra in both the Faraday and Voight geometries. This arrangement places the sample in the magnet outer vacuum chamber, which effectively provides a cold trap and requires one less window than the traditional variable temperature insert configuration.\cite{Padilla-magnet-RSI2004} Furthermore, we can take advantage of \emph{in situ} gold coating because the sample is in vacuum. This latter procedure permits the determination of absolute reflectance in magnetic field and corrects for any spurious effects due to field-induced misalignment.\cite{Homes-App-Optics1993} 

Reflectance and transmission spectra were recorded at selected temperatures from 6-295 K and magnetic fields 0-8 T, over the frequency range 30-6000 cm$^{-1}$. For photon energies above the mid-infrared the reflectance data were calculated from visible-UV ellipsometry up to 48 000 cm$^{-1}$.  Magnetoreflectance experiments were performed by first measuring the sample reflectance relative to a reference mirror for all temperatures and magnetic fields of interest. Then the sample was coated \emph{in situ} with gold and the entire measurement was repeated. The data were augmented with appropriate low- and high-frequency extrapolations and transformed via the Kramers-Kronig (KK) relations to obtain the frequency-dependent optical constants, including the optical conductivity $\hat{\sigma}(\omega)=\sigma_1(\omega)+i\sigma_2(\omega)$ The ellipsometry technique provides a direct, model-independent measurement of  the high-frequency optical constants, and serves as a constraint for extrapolations used in the KK calculation.
 
\subsection{Transmission experiment}

Transmission data for Bi$_2$Se$_3$ in zero magnetic field are displayed in Fig. 2(a). The thin crystal is transparent only within a narrow band in the mid-IR extending from 470-2000 cm$^{-1}$ at room temperature. The low-frequency absorption is indicative of free-carrier transport, while the high-frequency cutoff identifies the energy of direct allowed transitions from the valence band to the Fermi level. The onset of this later process at $\omega = \omega_g$ overestimates the energy $E_0$ of the gap between the valence and conduction band. The magnitude of $E_0$ can be determined by measuring a series of samples with varying carrier concentration and thus varying Fermi energy.\cite{Gorbrecht-gap-ZPhys1964} This interpretation of the transmission data is supported by $R(\omega)$ measurements and $\hat{\sigma}(\omega)$ results. At lower temperatures the low-frequency transmission onset becomes very steep and the mid-IR absorption edge moves to higher frequencies, partly due to a reduction in thermal level broadening. At 6 K the gap edge lies at $\omega_{g}$ = 2650 cm$^{-1}$ = 0.329 eV. Application of magnetic field perpendicular to the trigonal $c$ axis leads to a minor reduction in $\omega_{g}$. As seen in the transmission ratios $T(6$ K$, H)/T(6 $ K$, 0 $ T) in Fig. 2(b), transmittance decreases with field on both ends of the transmission window, particularly on the high-frequency side. The inset to Fig. 2(b) records the value of the direct transition energy at magnetic fields up to 8 T; a total shift of 3 cm$^{-1}$ is induced. Fields $H \parallel c$ initiate a more modest change in $\omega_{g}$.

\begin{figure}
\centering
   \includegraphics[width=3.375in]{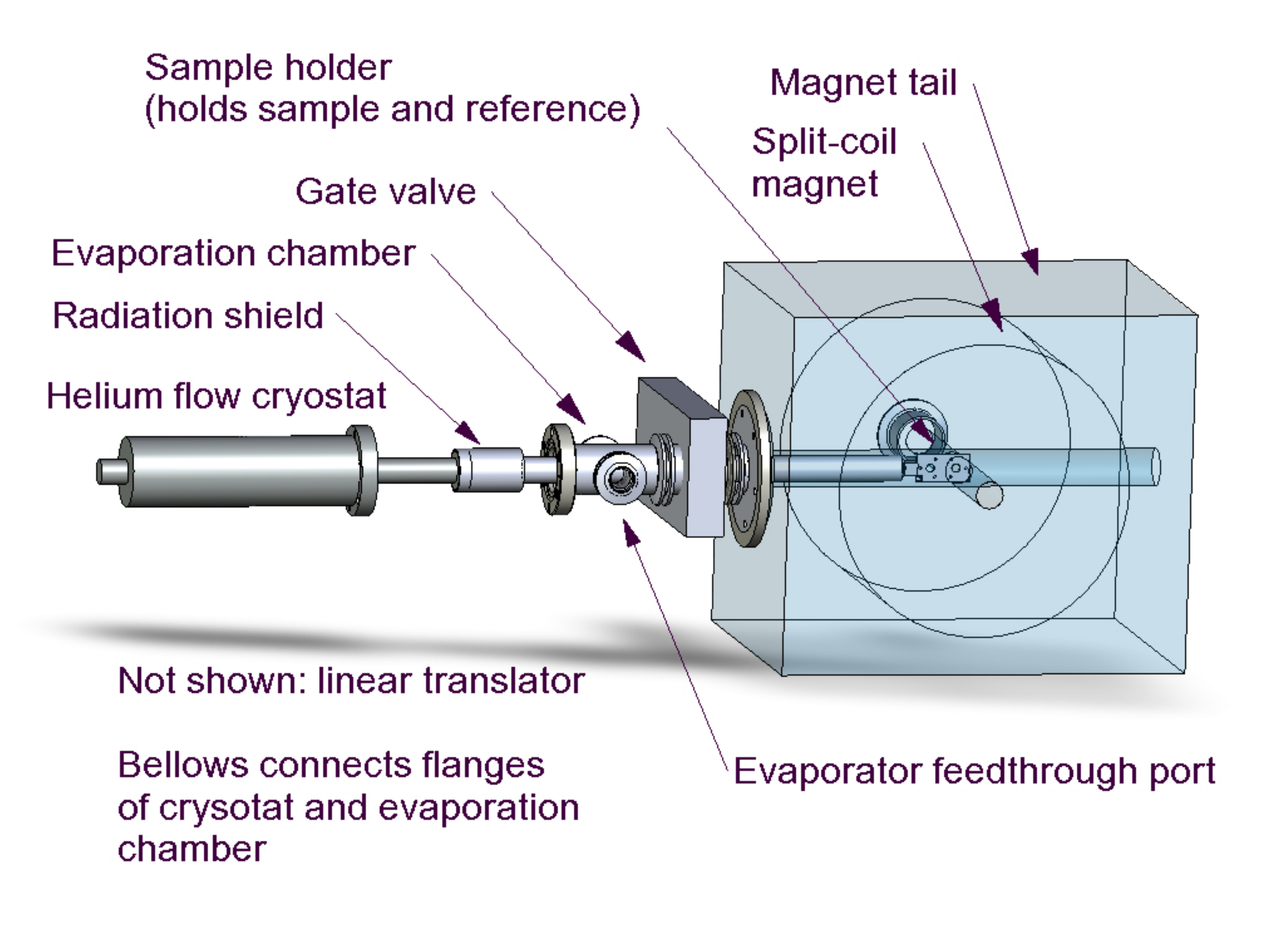}
     \caption{Novel magneto-spectroscopic apparatus designed and constructed at UCSD. A helium-flow crysostat with long copper extension cools the sample, which resides in the magnet outer vacuum chamber. A linear translator provides actuation between sample and reference at the focus of incident light inside the magnet. The sample may also be retracted to the evaporator port where gold coating is performed in vacuum. This arrangement allows for fully-automated collection of absolute reflectance data in magnetic field.} 
     \label{exp}
\end{figure}

\begin{figure}
\centering
   \includegraphics[width=3.375in]{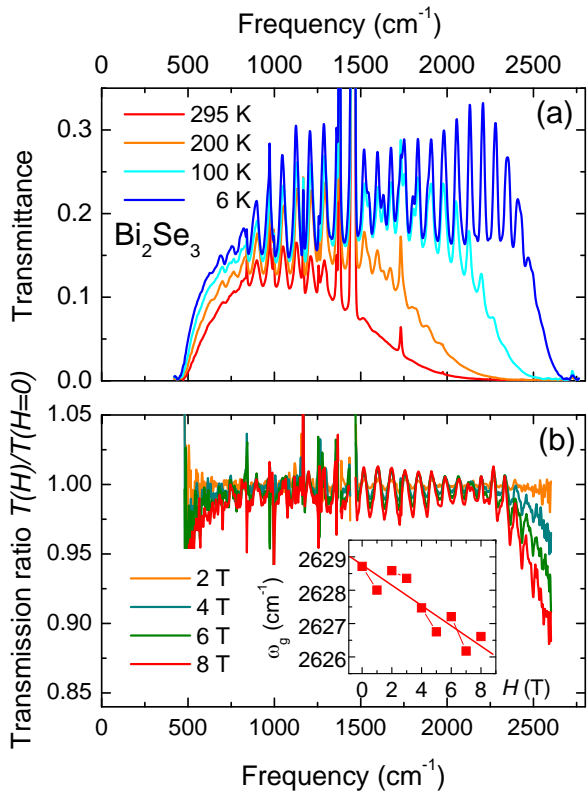}
     \caption{(a) Infrared transmittance of Bi$_2$Se$_3$ for a range of temperatures $T$ = 6-295 K. Sharp oscillatory features are Fabry-Perot interference fringes due to internal reflections in the thin sample. (b) Transmission ratios $T(6 $ K$, H)/T(6 $ K$, 0 $ T). Inset: suppression of direct allowed transition absorption edge $\omega_{g}$ by magnetic field $H \perp c$ at $T$ = 6 K.} 
     \label{trans}
\end{figure}

\subsection{Reflectance and optical conductivity in zero magnetic field}

Reflectance data for Bi$_2$Se$_3$ in zero magnetic field are found in Fig. 3. At room temperature the spectrum is dominated by a plasma edge in the far-IR, corroborating the picture of free electron transport deduced from the transmission.\cite{Drude-footnote} As seen in Fig. 3(a), the plasma edge hardens upon cooling to 200 K, then moves to lower frequencies and sharpens as the temperature is decreased to 6 K. Phonon absorption features are observed near 61 cm$^{-1}$ ($\alpha$ mode) and 133 cm$^{-1}$ ($\beta$ mode); both sharpen considerably at lower temperatures. These features are associated with relative transverse motion of charged Bi and Se ions (modes of $E_u$ symmetry)\cite{Richter-Kohler-Becker-Bi2Se3-phonons-PhysSolStat1977} and are displayed in more detail in Fig. 4(a). In the mid-IR near $\omega_B$ = 2400 cm$^{-1}$, a peak appears in the reflectance at low temperatures [Fig. 3(b)]. This phenomenon, known as the Burstein shift, marks the onset of direct, allowed interband transitions.\cite{Kohler-burstein-shift-SolStatPhys1974} A comparison of this peak frequency ($\omega_B$) with the energy for direct transitions determined from transmission ($\omega_g$)  yields an agreement within roughly 10\% (see Fig. 5). Reflectance data for the extended frequency range are shown in the inset to Fig. 3(b). 

\begin{figure}
\centering
   \includegraphics[width=3.375in]{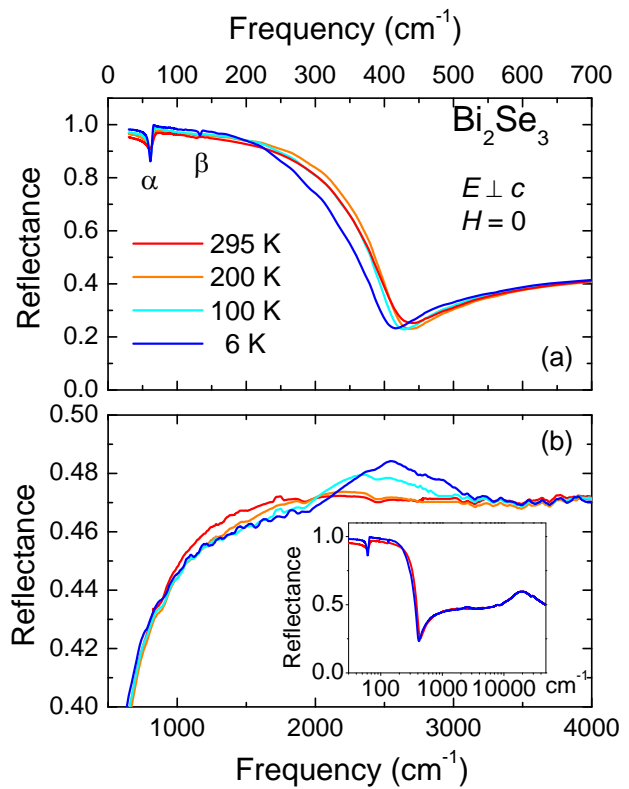}
     \caption{Infrared reflectivity of Bi$_2$Se$_3$ in zero magnetic field. The far-IR spectra (a) are characterized by a free-carrier plasma edge, as well as several phonon features. In the mid-IR range at low temperatures a peak develops near the direct transition energy. Inset: Reflectance over the extended frequency range. Reflectance above 5 000 cm$^{-1}$ was calculated from ellipsometric data.} 
     \label{ZFrefl}
\end{figure}

\begin{figure}
\centering
   \includegraphics[width=3.375in]{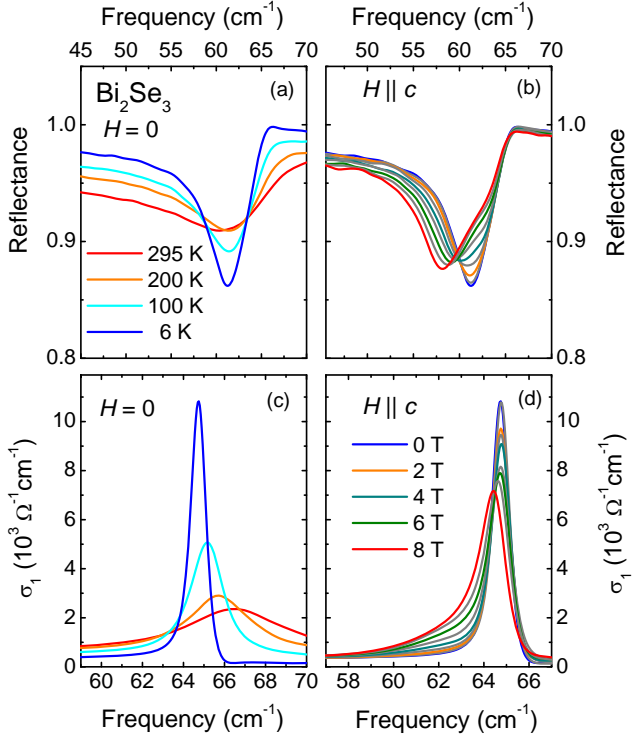}
     \caption{Infrared reflectance (top panels) and optical conductivity (bottom panels) for temperatures 6-295 K (left panels) and magnetic fields parallel to the $c$ axis (right panels). The anomalously strong phonon is softened and asymmetrically broadened in magnetic field $H \parallel c$. For $H \perp c$ no softening or broadening of this mode is observed.} 
     \label{phonon}
\end{figure}

\begin{figure}
\centering
   \includegraphics[width=3.375in]{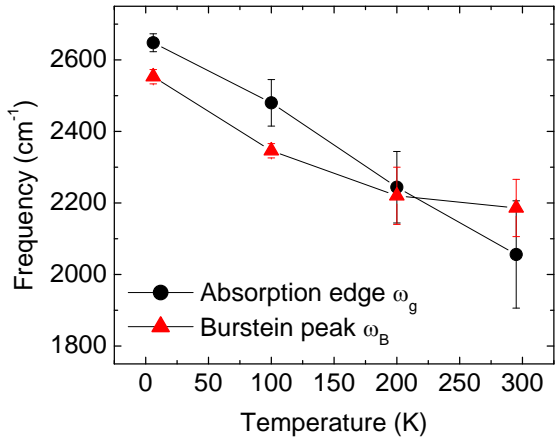}
     \caption{Mid-IR absorption edge ($\omega_g$, from transmittance) and Burstein peak ($\omega_B$, from reflectance) as a function of temperature.} 
     \label{BandgapBurstein}
\end{figure}

Figure 6(a) displays the real part of the optical conductivity $\sigma_1(\omega)$, obtained by Kramers-Kronig transformation of the reflectance. The free-carrier response is plainly visible as a Drude oscillator (Lorentzian centered at zero frequency) which sharpens considerably with decreasing temperature.  An oscillator fit reveals a collapse in the quasiparticle scattering rate, which experiences a threefold decrease, from 60 $\rightarrow$ 22 cm$^{-1}$ at low temperature.\cite{scattering-rate-footnote} This sharpening uncovers a broad electronic mode (labeled here as $\gamma$) in the far-IR centered at $\omega_{\gamma}$ = 300 cm$^{-1}$. The phonons also sharpen [Fig. 4(c)], as well as experience a shift of eigenfrequency: $\omega_{\beta}$ increases by 5 cm$^{-1}$, while  $\omega_{\alpha}$ decreases by 2 cm$^{-1}$.  The $\alpha$ mode, which has previously been identified as having an oscillator strength much greater than theory would predict,\cite{Richter-Kohler-Becker-Bi2Se3-phonons-PhysSolStat1977} exhibits an asymmetric Fano lineshape, and will be discussed in more detail in Sec. IIIB below. At higher frequencies, the Burstein shift causes a significant transfer of spectral weight from a gapped region spanning $1000 < \omega < 2500$ cm$^{-1}$ to a peak at 3000 cm$^{-1}$. The direct gap at 2500 cm$^{-1}$ produces a relatively weak feature in the conductivity. The largest feature in the conductivity is the triplet of interband absorptions centered at $\omega$ = 16 000 cm$^{-1}$. These interband features are not the focus of this work, but are discussed in detail for the related compound Bi$_2$Te$_3$ in Ref. \onlinecite{Greenaway-Harbeke-Bi2Se3-band-structure-JPCS1965}.

\begin{figure}
\centering
   \includegraphics[width=3.375in]{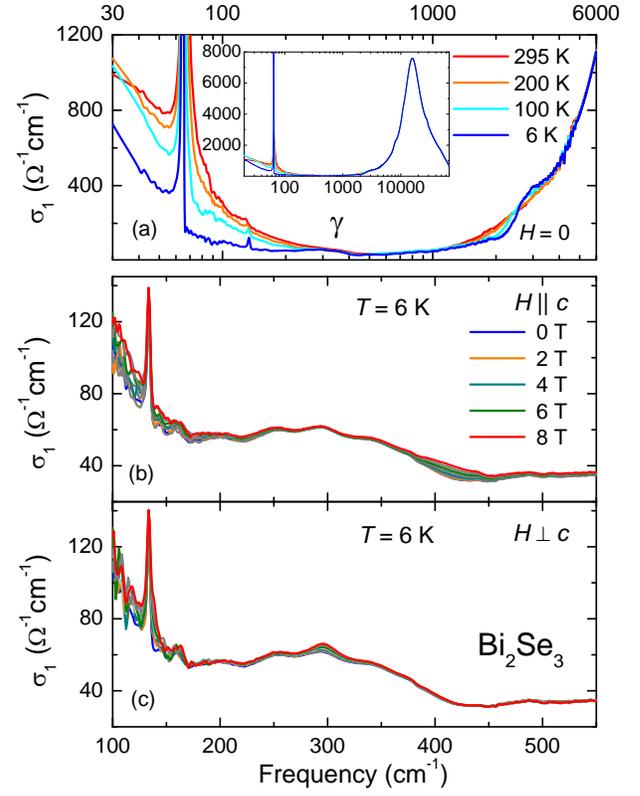}
     \caption{Optical conductivity $\sigma_1(\omega)$ for Bi$_2$Se$_3$ in zero magnetic field (a), as well for magnetic fields applied parallel (b) and perpendicular (c) to the trigonal $c$ axis. Inset: Optical conductivity over a wide frequency range.} 
     \label{Sig1}
\end{figure}

\subsection{Reflectance and optical conductivity in magnetic field}
The application of magnetic field $H \parallel c$ initiates several distinct effects in the infrared response. Looking first at the reflectivity in Figs. 4(b) and  7(a), we see that the absorption feature corresponding to the $\alpha$ phonon undergoes both a shift to lower frequencies and a change in lineshape. At the same time, the reflectance in the plasma minimum is enhanced without changing the overall lineshape of the plasmon. The $\beta$ mode is not modified by $H \parallel c$, and no measurable change of reflectance occurs in the mid-IR. The optical conductivity spectra in field [Figs. 4(d) and 5(b)] exhibit an overall broadening of the Drude peak and $\gamma$ mode. The Drude broadening is consistent with reports of positive magnetoresistance.\cite{Kulbachinskii-BiSbSe-SdH-PRB} The spectra also illustrate more clearly the shift of the $\alpha$ phonon to lower frequencies and reveal that the mode is being asymmetrically broadened. 

\begin{figure}
\centering
   \includegraphics[width=3.375in]{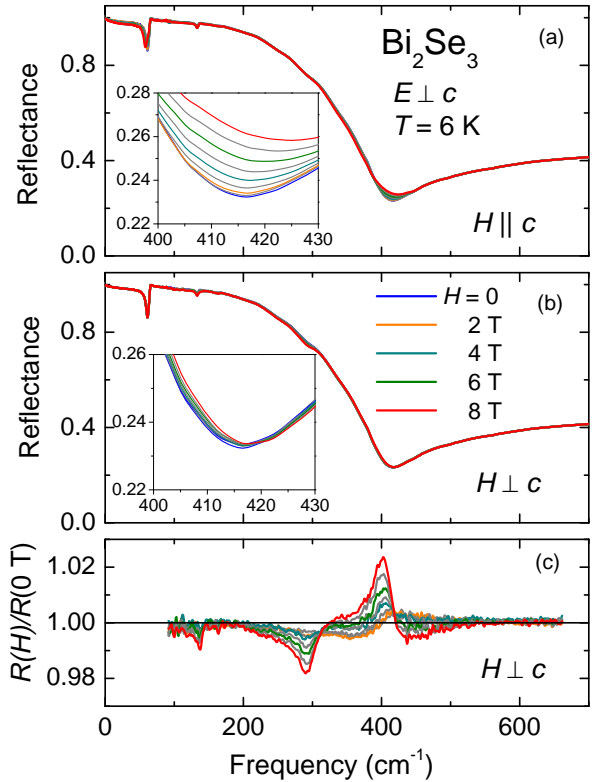}
     \caption{Infrared reflectivity of Bi$_2$Se$_3$ for magnetic fields applied parallel (a) and perpendicular (b) to the trigonal $c$ axis. Also plotted (c) are the reflectance ratios $R(6 $K$, H)/R(6 $K$, 0 $T) } 
     \label{HRefl}
\end{figure}

For the other relevant magnetic field orientation, $H \perp c$, the changes in reflectance are much subtler. For this reason we show in Fig. 7 not only the absolute reflectance [Fig. 7(b)], but also the magnetic field ratios $R(6 $ K$, H)/R(6 $ K$, 0 $ T) [Fig. 7(c)]. Viewing the ratios, it is clear that two distinct changes are promoted by the field: a reflectance dip at $\omega \approx 390$ cm$^{-1}$ and a shift of the plasma edge to higher frequency. The latter effect is shown in more detail in the inset to Fig. 7(b), and constitutes a total shift of $\Delta \omega_p \approx 1.5$ cm$^{-1}$. If we turn to $\sigma_1(\omega)$, shown in Fig. 6(c), it is apparent that the lower-frequency dip seen in $R(\omega)$ corresponds to an increase in the oscillator strength of the electronic $\gamma$ mode. This is in contrast to $H \parallel c$, which only increased the width of the $\gamma$ mode. The $\alpha$ phonon mode experiences a minor reduction in oscillator strength, but no shift in frequency.

\section{ Analysis and discussion}

\subsection{Integrated spectral weight}

A comprehensive understanding of the evolution of the optical conductivity with temperature and magnetic field can be gained through examination of the integrated spectral weight $N(\omega) =\int_{0}^{\omega} d\nu ~\sigma_{1}(\nu)$. $N(\omega)$ quantifies the oscillator strength of excitations occurring below frequency $\omega$, and is useful for assessing the energy scales involved in electronic processes. Figure 8(a) illustrates the zero-field transfer of spectral weight relative to the value at room temperature spectra as $N(\omega, T)/N(\omega,$ 295 K). At $T$ =  200 K, the narrowing of the Drude mode is accompanied by a modifcation of spectral weight over the entire far-IR, until the room-temperature value is reached near $\omega$ = 1000 cm$^{-1}$. For $T \le 200 $, the low-frequency spectral weight is conserved below a scale of approximately $3 \times \frac{1}{\tau}$, as evidenced by the convergence of the curves at that energy. At $T$ = 6 K the gap in $\sigma_1(\omega)$ results in a redistribution of $N(\omega)$ to energies exceeding the band gap (2500 cm$^{-1}$), until conservation of spectral weight is reached near $\omega$ = 4000 cm$^{-1}$.

In the presence of an applied magnetic field most changes in spectral weight involve a reshuffling of weight between the Drude, phonon, and $\gamma$ modes. For $H \parallel c$ we observe in Fig. 8(b) a broadening of the Drude peak which is fully contained within 250 cm$^{-1}$. $N(\omega)$ is equal for all magnetic field values from 250-350 cm$^{-1}$, above which $N(\omega, 8 $ T) increases slightly due to the high-frequency broadening of the $\gamma$ mode. This anomalous spectral weight appears to have been transferred from the higher-energy portion of the spectrum, originating from small changes in reflectance spread over a broad mid-IR frequency range. This phenomenon is known to occur in other materials, including the cuprate superconductors.\cite{LaForge-sumrule-PRL2008,LaForge-sumrule-PRB2009} In fields oriented perpendicular to the $c$ axis [Fig. 8(c)], very little modification of the Drude conductivity is observed. Instead, a small transfer of spectral weight occurs between the $\alpha$ phonon and the $\gamma$ mode. This weight is largely regained by 500 cm$^{-1}$, but may extend slightly higher in frequency. 

\begin{figure}
\centering
   \includegraphics[width=3.375in]{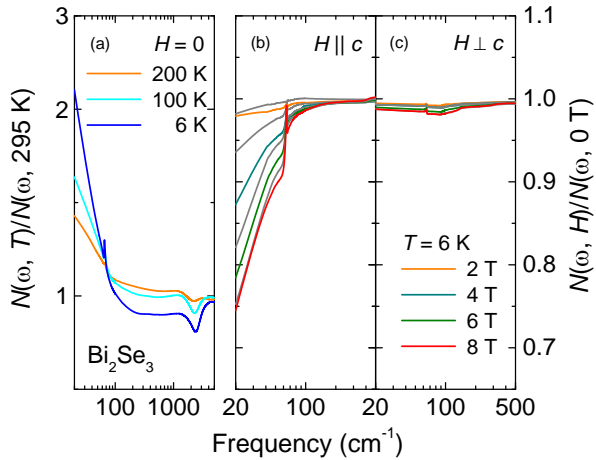}
     \caption{(a) Integrated spectral weight $N(\omega) = \int_{0}^{\omega} d\nu ~\sigma_{1}(\nu)$ at several temperatures in zero magnetic field. Spectral weight is redistributed over an energy range as high as 4000 cm$^{-1}$. Spectral weight magnetic field ratios $N(\omega, H)/N(\omega, $0 T) are also shown for magnetic fields oriented parallel (b) and perpendicular (c) to the $c$ axis. Significant redistribution of spectral weight across the far-IR is observed for $H \parallel c$, while very little change occurs for $H \perp c$.} 
     \label{SW}
\end{figure}

\subsection{Electron-phonon coupling and magnetoelectric effects}
It was noted in a previous optical study of the related compound Bi$_2$Te$_3$ (Ref. \onlinecite{Richter-Kohler-Becker-Bi2Se3-phonons-PhysSolStat1977}) that the lowest-frequency infrared active phonon, analogous to the $\alpha$ mode at 64 cm$^{-1}$ in Bi$_2$Se$_3$, displays an oscillator strength 60\% greater than that predicted from a Born-van Karman lattice model. From this it was concluded that polarization effects must play a large role in the movements of the VI$^{(2)}$ atoms. Furthermore, it is clear from examination of the conductivity spectra  in Fig. 4(d) that the phonon lineshape is strongly asymmetric, a signature of electron-phonon coupling and Fano physics.  The Fano lineshape, ubiquitous across the  branches of physics, appears in systems in which a discrete mode is coupled to a continuum of excitations.\cite{Fano-PR1961} Quantum interference between the wave functions for two transition pathways between the ground state and an excited state (one directly to a discrete excited state; the other to the continuum, and then to the discrete state) results in an asymmetric scattering cross section. This phenomenon is a common indicator of hybridization between phonons and conduction electrons in materials with high polarizability. 

To better understand this behavior, we have applied to the data a fitting analysis which models the $\alpha$ phonon as a Fano resonance of the form 

\begin{equation}
\sigma_{1, Fano}(\omega)=\frac{\omega_{p,Fano}^2}{4 \pi \Gamma} \frac{q^2 +2q\epsilon-1}{q^2(1+\epsilon^2)},
\end{equation}
as a function of the reduced energy $\epsilon=(\omega-\omega_{\alpha})/(\Gamma)$.
 This lineshape is characterized by the center frequency $\omega_{\alpha}$, linewidth $\Gamma$, and plasma frequency $\omega_{p,Fano}$, similar to a Lorentz oscillator. The Fano parameter $q$ determines the asymmetry of the resonance, and is a useful measure of the degree of coupling between the discrete and continuum modes.\cite{Abstreiter-Cardona-scattering-1984} The Lorentz lineshape is recovered for $|q| \rightarrow \infty $ and the sign of $q$ sets the direction of asymmetry.  All other spectral features, including free-carrier and interband absorptions, were modeled with classical Lorentz oscillators. A representative fit for $H = 4$ T is compared to the far-IR experimental reflectance data in Fig. 9(a). The measured lineshape is clearly reproduced by the Fano functional form. Fitting parameters for  the $\alpha$ phonon and Drude peak are displayed in Table I. 

The evolution of the fit parameters for the $\alpha$ and Drude modes with magnetic field yields insight into the nature of free electron-lattice coupling in Bi$_2$Se$_3$. Since changes to the $\alpha$ mode are minimal for $H \perp c$, we will focus on data for $H \parallel c$. The most prominent overall effect of magnetic field upon the conductivity spectra in Fig. 6(b) is the increase in the linewidth $1/\tau^{Drude}$ of the Drude peak. This parameter has been extracted from the oscillator fit and is plotted with triangles in Fig. 9(b). We see that $1/\tau^{Drude}$ is constant at low fields, increases sharply  for intermediate fields, and saturates at 30 cm$^{-1}$ at $H$ = 7 T.  The width of the modified phonon (not shown) increases linearly with field over entire range. The asymmetry parameter $q$, indicated by squares in Fig. 9(b), scales with $1/\tau^{Drude}$, including the same plateau features for $H < 3$ T and $H >$ 7  T. This correlation suggests that the strength of coupling between the two modes may increase as the free-carrier scattering frequency approaches that of the lattice vibration. The negative value of $q$ indicates that the phonon is primarily interacting with electronic absorption at $\omega > \omega_{\alpha}$.\cite{Damascelli-FeSi-PRB1997,Damascelli-diss1999}  Interactions between the $\alpha$ mode and $\gamma$ mode may exist as well, since the latter is strongly field dependent as well. 

Magnetic field-induced modification of the Fano effect has been observed in several areas of condensed matter physics; the role of the field in modifying the response is unique to each system. In experiments involving mesoscopic coupled-quantum-dot interferometers\cite{Kobayashi-fano-quantum-dot-PRB004} and coupled carbon nanotubes,\cite{Kim-nanotubes-fano-Hfield-PRL2003} the magnetic field modifies the Fano interaction via the Aharonov-Bohm effect, adding an arbitrary phase to one of the transition pathways. In ultrathin epitaxial semiconductors, application of magnetic field tends to symmeterize Fano resonance features due to Landau-level confinement.\cite{Vasilenko-fano-thin-GaAs-Semicond1999} This is in contrast to the situation in bulk intrinsic semiconductors, where Fano resonances are created when magnetoexcitons overlap in energy with continuum band states.\cite{Glutsch-Chemla-fano-bulk-GaAs-PRB1994}

The phonon softening in magnetic field which we observe is, to the best of our knowledge, unprecedented for a non-magnetic system that does not undergo a phase transition. At this point we speculate that the field-induced phenomena of phonon asymmeterization and band-gap edge shifting might be related through the mechanism of magnetostriction. Mechanical deformation with applied magnetic field is common in systems with enhanced spin-orbit coupling,\cite{Goodenough-spin-orbit-magnetostriction-PRB1967} a defining characteristic of the topological insulator class of materials. Magnetostrictive behavior can also be found in more similar systems, such as elemental bismuth.\cite{Michenaud-Bi-magnetostriction-PRB1982} In this latter case the many-valley nature of the band structure leads to bands with differing magnetization energies. In a magnetic field, electron-transfer processes redistribute carriers to minimize the free energy and a strain is induced. Given that Bi$_2$Se$_3$ shares this many-valley form of the band structure, a similar mechanism might provide a cohesive explanation for the simultaneous change in lattice dynamics and shift of the band edge with magnetic field. Further measurements are necessary to determine if appreciable magnetostriction occurs in Bi$_2$Se$_3$. 

Magnetoelectric coupling uncovered by our experiments could further be related to the topological nature of the material. The topological magnetoelectric effect, in which an applied magnetic (electric) field creates an electric (magnetic) field in the same direction, is predicted to occur in a nontrivial topological insulator.\cite{Qi-Zhang-topological-field-theory-PRB2008,Essin-Moore-TME-PRL2009} The sensitivity of infrared active phonon modes to perturbations of local electric fields is well-known. It has been used to explore collective electronic modes in cuprate superconductors\cite{Munzar-local-field-SolStatCom112-1999} and dielectric effects for insulators in applied electric field,\cite{Li-Basov-TiO2-electric-field-APL2005} to name several examples. Based on the dramatic changes to phonon modes we observe in Bi$_2$Se$_3$, it is clear that the local electric fields acting on the Bi ions are modified by magnetic field. This behavior is consistent with the expected topological magnetoelectric polarization. The fact that magnetoelectric effects are only observed for the $\alpha$ mode when the magnetic field is applied in the direction of  the bismuth displacement provides further support for the influence of spin-orbit coupling. Nontrivial contributions to the electromagnetic response may be augmented by topological states available in the interior of the crystal. Here defects may reduce coupling between quintuple layers and effect  local decoupling.

\begin{figure}\centering
   \includegraphics[width=3.375in]{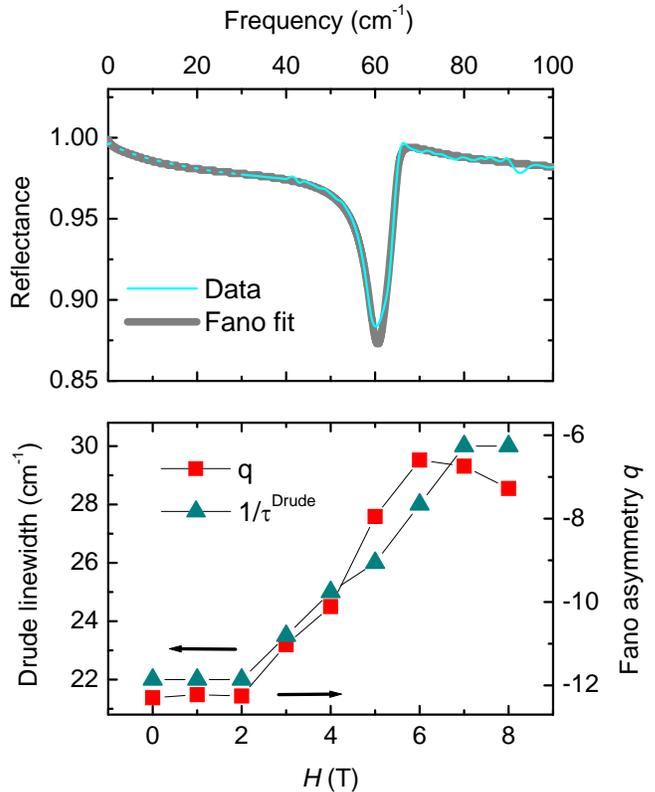}
     \caption{(a) Representative fit (thick line) of Fano lineshape to far-IR reflectance data (thin line). Here, $T$ = 6 K and  $H$ = 4 T is applied parallel to the $c$ axis. (b) Fit parameters $q$ (square symbols) and ${1/\tau} ^{Drude}$ (triangle symbols) as a function of magnetic field. The Fano asymmetry scales with the linewidth of the Drude conductivity.}
     \label{Fano-fit}
\end{figure}

\bigskip

\begin{table*}{}
\caption{Parameters used for low-frequency oscillators in Fano/Lorentz fit to infrared spectra at $T$ = 6 K, $H \parallel c$. A representative fit is shown in Fig. 9. }
\renewcommand{\arraystretch}{1.5}
\begin{tabular*}{5.5in}{l @{\extracolsep{\fill}}c c c c c c}
\hline
\hline
$H$ (T) & $\omega_{p,Drude}^2$ (10$^6$ cm$^{-2}$) & $1/\tau_{Drude}$ (cm$^{-1})$ & $\omega_{\alpha}$ (cm$^{-1}$) & $\omega_{p,Fano}^2$ (10$^5$ cm$^{-2}$)  & $\Gamma$ (cm$^{-1}$) & $q$\\ \hline
0 & 2.52 & 22 & 64.77 & 5.07 & 0.76 & -12.3\\ 
1 & 2.52 & 22 & 64.82 & 5.16 & 0.77 & -12.2\\ 
2 & 2.52 & 22 & 64.80 & 4.99 & 0.84 & -12.3\\ 
3 & 2.63& 24 & 64.81 & 5.03 & 0.88 & -11.0\\ 
4 & 2.89 & 25 & 64.83 & 5.00 & 0.91 & -10.1\\ 
5 & 3.28 & 26 & 64.83 & 5.16 & 1.08 & -8.0\\ 
6 & 3.43 & 28 & 64.80 & 5.27 & 1.14 & -6.6\\ 
7 & 3.43 & 30 & 64.70 & 5.60 & 1.28 & -6.7\\ 
8 & 3.33 & 30 & 64.48 & 5.87 & 1.42 & -7.3\\ 
\hline
\hline
\end{tabular*}

\end{table*}

\bigskip

\subsection{Assessing relevant magnetic field scales via comparsion to elemental bismuth}

Changes of the optical spectra of Bi$_2$Se$_3$ with magnetic field, while informative, are fairly mild. This is in part due to the small magnitude of the experimental magnetic field relative to the field scales of the electronic system. According to magneto-transport measurements at low temperatures, magnetic fields on the order of 16 T, twice those accessible in the present study, are sometimes needed to resolve Shubinikov-de Haas (SdH) oscillations of the Hall coefficient.\cite{Kulbachinskii-BiSbSe-SdH-PRB} Recently, however, SdH oscillations have been observed at much lower magnetic fields (approximately 3 T).\cite{Analytis-Shen-Bi2Se3-SdH-ARPES-condmat2010} Cyclotron absorption should also conceivably be observed at lower fields.  Given the effective mass determined for $H \parallel c$ from magnetoabsorption ($m^*=0.105 m_e$)\cite{Kulbachinskii-BiSbSe-SdH-PRB} or ARPES/SdH [$m^*=(0.125-0.15) m_e$] measurements,\cite{Analytis-Shen-Bi2Se3-SdH-ARPES-condmat2010} the cyclotron frequency should be $\omega_c = e H/{2 \pi c m^*}$ = 50-70 cm$^{-1}$ at 8 T. Since the free-carrier scattering rate is approximately $1/\tau^{Drude}= 30$ cm$^{-1}$, the Landau-level effects should be within the experimental window, yet we do not observe them in our experiment.

This behavior can be contrasted with that of elemental bismuth, which has much lighter charge carriers ($m^*  = 0.004$) and lower carrier density, rendering it more susceptible to the magnetic field. Indeed, in Bi quantum oscillations have been observed at fields less than 1 T, and the quantum limit, the magnetic field at which all carriers reside in the lowest Landau levels, is only 9 T. Figure 10 displays reflectance data for Bi at $T$ = 6 K in magnetic fields up to 8 T, just below the quantum limit. Here we see a dramatic reconstruction of the electromagnetic response by magnetic field: the sharp plasma edge at 160 cm$^{-1}$ in zero field has become nearly unrecognizable, and the low-frequency reflectance is strongly suppressed. Analysis and comparison of the optical conductivity are more complicated because the small effective mass precludes the use of the KK transformation in high magnetic field. Appropriate magneto-optical fitting analyses have been successfully carried out in similar systems such as graphite.\cite{Li-Basov-graphite-magneto-optics-PRB2006}

\begin{figure}
\centering
   \includegraphics[width=3.375in]{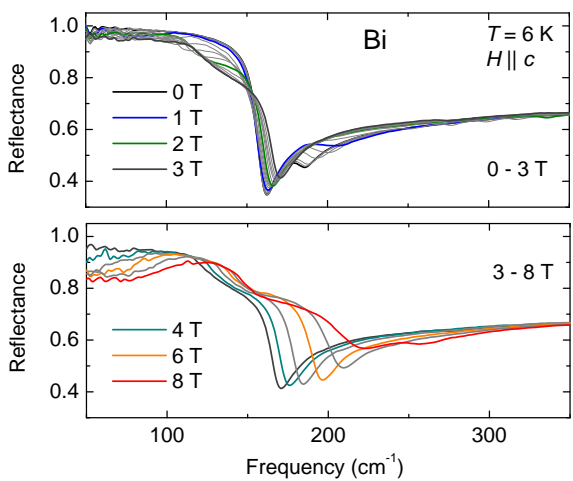}
     \caption{Infrared reflectance of elemental bismuth in magnetic fields oriented parallel to the trigonal $c$ axis. Maximum measured field is just below the quantum limit of 9 T. } 
     \label{Bi}
\end{figure}

\section{Summary and outlook}
We have determined the optical conductivity $\hat{\sigma}(\omega)$ in magnetic field for the narrow-gap semiconductor Bi$_2$Se$_3$ through a combination of infrared reflectance and UV-visible ellipsometry. A gap in the conductivity induces an extensive redistribution of spectral weight over the mid-IR frequency region. Magnetic fields $H \perp c$ enhance the broad electronic absorption in the far-IR, as well as slightly shift the metallic plasma edge minimum. For $H \parallel c$, we have discovered magnetoelectric coupling by monitoring phonon response: a strong low-frequency phonon is significantly modified by the field and a Fano lineshape analysis reveals an interesting scaling between the phonon asymmetry and the broadening of the Drude peak. 

These observations present two challenges for the understanding of Bi$_2$Se$_3$. First, the phonon effects observed in magnetic field are highly anomalous, and unprecedented for a system such as  Bi$_2$Se$_3$, which is non-magnetic and without phase transitions. Such behavior necessitates further investigation of the interplay of spin-orbit effects prominent in Bi and Bi-based compounds with phonons that are sensitive to local electric fields. This represents an alternative approach for monitoring local magnetoelectric effects. Second, the absence of cyclotron resonance effects is highly unusual. Based on effective mass values determined in magneto-oscillation experiments,\cite{Kulbachinskii-BiSbSe-SdH-PRB,Analytis-Shen-Bi2Se3-SdH-ARPES-condmat2010} cyclotron resonance effects are expected to be visible in our experimental window.  Future experimental studies will investigate crystals which are closer to the stoichiometric levels in an attempt to optically probe other expected consequences of the topological insulator state. 

\section*{Acknowledgments}
Research at UCSD was supported by DOE-BES. The authors gratefully acknowledge fruitful discussions with Shoucheng Zhang.

\end{document}